\documentclass[twocolumn,showpacs,preprintnumbers,amsmath,amssymb]{revtex4}

\usepackage{graphicx}
\usepackage{amssymb}
\usepackage{dcolumn}

\newcommand {\be}{\begin{equation}}
\newcommand {\ee}{\end{equation}}
\newcommand {\bea}{\begin{eqnarray}}
\newcommand {\eea}{\end{eqnarray}}
\newcommand {\FIG}[1]{Fig. \ref{#1}}
\newcommand {\EQ}[1]{Eq. (\ref{#1})}

\newcommand {\PRA}[1]{Phys. Rev. A {\bf {#1}}}

\newcommand {\PRL}[1]{Phys. Rev. Lett. {\bf {#1}}}

\newcommand {\NAT}[1]{Nature {\bf {#1}}}

\begin{document}
\title{Random walks and diameter of finite scale-free networks}

\author{Sungmin Lee}
\author{Soon-Hyung Yook}
\email{syook@khu.ac.kr}
\author{Yup Kim}
\email{ykim@khu.ac.kr}
\affiliation{Department of Physics and
Research Institute for Basic Sciences, Kyung Hee University, Seoul
130-701, Korea}
\date{\today}

\begin{abstract}

Dynamical scalings for the end-to-end distance $R_{ee}$ and the
number of distinct visited nodes $N_v$ of random walks (RWs) on
finite scale-free networks (SFNs) are studied numerically. $\left<
R_{ee} \right>$ shows the dynamical scaling behavior
$\left<R_{ee}({\bar \ell},t)\right>= \overline{\ell}^\alpha
(\gamma , N) g(t/\overline{\ell}^z )$, where $\overline{\ell}$ is
the average minimum distance between all possible pairs of nodes
in the network, $N$ is the number of nodes, $\gamma$ is the degree
exponent of the SFN and $t$ is the step number of RWs. Especially,
$\left<R_{ee}({\bar \ell},t)\right>$ in the limit $t \rightarrow
\infty$ satisfies the relation $\left< R_{ee} \right> \sim
\overline{\ell}^\alpha \sim d^\alpha$, where $d$ is the diameter
of network with $d ({\bar \ell}) \simeq \ln N$ for $\gamma \ge 3$
or $d  ({\bar \ell}) \simeq \ln \ln N$ for $\gamma < 3$. Based on
the scaling relation $\left< R_{ee} \right>$, we also find that
the scaling behavior of the diameter of networks can be measured
very efficiently by using RWs. 
\end{abstract}
\maketitle

For almost a decade there have been many studies on the
topological properties of complex networks, since many structures
of physically interacting systems are shown to form nontrivial
complex structures \cite{DM,Jeong01}. In these studies much effort
has been put to investigate the physical origin of complex
networks \cite{handbook_born, evolnet}. It has been found that
most of real web-like systems share several prominent structural
features: scale-free degree distribution, high clustering, and
small average path length (APL) \cite{handbook_born, evolnet, DM,
strog}. It is well known \cite{randgraph, majum} that random
networks (RNs) \cite{ER1} and small-world networks (SWNs)
\cite{wattsstrog} have very small APL or diameter $d$ which scales
as $d \sim \ln N$, where $N$ is the number of nodes. Recently,
Cohen and Havlin \cite{ultra} studied the diameter of scale-free
networks (SFNs) \cite{DM} whose degree distribution $P(k)$
satisfies a power law $P(k) \sim k^{-\gamma}$. Using analytical
arguments they showed \cite{ultra} that

\begin{equation}\label{d}
d (N) \sim \left\{
\begin{array}{lll}
\ln \ln N & , & 2< \gamma < 3\\
\ln N / \ln \ln N & , & \gamma = 3\\
\ln N & , & \gamma > 3.
\end{array}\right.
\end{equation}
For measuring the APL, the breath-first algorithm (BFA)
\cite{order} are mainly used. This algorithm is known to scale as
$O(N^2)$. In this paper, we will show that random walk (RW) can
provide more efficient method to find the scaling behavior of $d$
like Eq. (\ref{d}).

The early studies on complex networks mainly focused on their
topological properties. Recently, the physical systems whose
elements interact along the links in complex networks have drawn
much attention. Furthermore a number of studies have focused on
the effects of the underlying topologies on the dynamical
properties of such systems. Many dynamical systems on complex
networks show rich behaviors which are far from the mean-field
expectations and affected by the underlying topology
\cite{Critical, Percolation}. For example, the dynamical
properties of RWs on complex networks have been shown to be
closely related to the topology of underlying networks
\cite{Noh92, noh_tree_loop, almaas}. Especially, the average
number of distinct visited sites and the average end-to-end
distance of RWs on SWNs are known to satisfy the scaling law
\cite{almaas}
\begin{equation}
O(p,N,t)=O_{sat} F(p^2 t,pN)~.
\end{equation}
Here, $p$ is a density of shortcut and $O_{sat}$ is a saturated
quantity.

It is well known that the SFNs have heterogeneous structures in
which nodes with anomalously large number of degree and nodes with
small degree co-exist \cite{DM,Jeong01}. In SFNs, the dynamical
properties of several systems are affected by the second moment of
degree distribution $\left< k^2 \right>$ \cite{Ising, bsonsfn,
cmaonsfn, lambliononsfn}. It is interesting to study how the
structural heterogeneity affects the scaling properties of RWs as
changing the degree exponent $\gamma$. In this paper, we mainly
investigate the dynamical scaling relation for  the end-to-end
distance $R_{ee}$ of RWs on finite SFNs with various $\gamma$.
From the scaling relation, it will be  shown that RWs on SFNs can
provide much more efficient method to measure the scaling behavior
of the diameter of finite SFNs than BFA. For the complementary
purpose we also study the scaling relation for the number of
distinct visited nodes $N_v$.

All the models are defined on a graph $Gr=\{N,K\}$, where $N$ is
the number of nodes in the network and $K$ is the total number of
degrees with the average degree $\left<k\right>=2K/N$. Networks of
the sizes $N=10^3 \sim 10^6$ are used. To generate SFNs, we use
the static network model introduced by Ref. \cite{SNU_staticmodel}
with $\left<k\right>=4$. Initially a random walker is placed on a
randomly chosen node denoted by $s$. At the time step $t+1$, the
walker jumps to a randomly chosen node among the nearest neighbor
nodes of the node where the walker is at $t$. The probability
$P(i,t;s,0)$ to find the walker at node $i$ at $t$ thus follows
the relation
\begin{equation}
P(i,t+1;s,0)=\sum^N_{j=1} \frac{A_{ij}}{k_j} P(j,t;s,0)
\end{equation}
with the condition $P(i,0;s,0)=\delta_{i,s}$ \cite{noh_tree_loop}.
Here, $A_{ij}$ is the adjacency matrix whose elements are
$A_{ij}=1 (0)$ if two nodes $i$ and $j$ are connected
(disconnected). All quantities are averaged over $100$ network
realizations and $1000$ different initial positions of RWs for
each network realization.

First, let us discuss the average end-to-end distance
$\left<R_{ee}\right>$ of RW on SFNs. At each time step $t$, we
measure the shortest distance, $R_{ee}(t)$, from the node where
the random walker is to the node $s$. By averaging $R_{ee}(t)$
over different initial positions and network realizations we get
$\left<R_{ee}\right>$. Here the shortest distance between two
nodes in networks means the shortest path length or the minimal
number of steps between them.

On a $D$-dimensional regular lattice, the average end-to-end
distance follows $\left<R_{ee}\right> \sim t^\nu$ with $\nu=1/2$
\cite{rrwalk13}. However, on finite systems, $\left<R_{ee}\right>$
saturates to a finite value in the limit $t \rightarrow \infty$.
In this limit, the probability for each site to be occupied by a
random walker is the same for all the sites and thus
$\left<R_{ee}\right> \sim \overline{\ell}$, where
$\overline{\ell}$ is the minimum distance averaged over all
possible pairs of sites \cite{almaas}. On a SWN, for small $t$
($\ll 1/p^2$) regime $\left<R_{ee}\right>$ again satisfies the
same relation as on the regular lattice ($\left<R_{ee}\right> \sim
t^{1/2}$), because the random walker cannot actually see
shortcuts. But in the large $t$ regime ($t \gg
\sqrt{\overline{\ell^2}}/p$), $\left<R_{ee}\right> \sim
\overline{\ell}$ . For the intermediate time region $1/p^2 \ll t
\ll \sqrt{\overline{\ell^2}}/p$, Almaas et al. expected that
$\left<R_{ee}\right> \sim \overline{\ell}$ satisfies the relation
$\left<R_{ee}\right> \sim t p$ \cite{almaas}.

\begin{figure}[h]
\includegraphics[width=7cm]{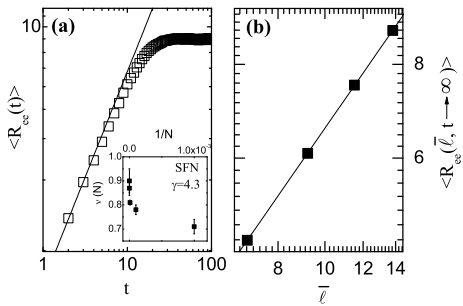}
\caption{Average end-to-end distance, $\left<R_{ee}\right>$ on
SFNs with $\gamma=2.4$ for $N=10^7$. (a) Plot of
$\left<R_{ee}\right>$ against $t$ in log-log scale. The solid line
correspond to $\nu=0.90$. The inset displays the dependence of the
measured $\nu$ on $1/N$. From the inset, we expect that
$\nu\rightarrow 1$ in the limit $N\rightarrow \infty$. (b) Plot of
$\left<R_{ee}({\bar \ell},t\rightarrow \infty)\right>$ against
$\bar\ell$ in log-log scale with ${\bar\ell}\sim \ln N$. The solid
line corresponds to $\alpha=0.93$.}\label{fignu}
\end{figure}

A typical dependence of $\left<R_{ee}\right>$ on $t$ for finite
SFNs is shown in \FIG{fignu}.
Figure 1 shows the measured $\left<R_{ee}\right>$ on SFNs with
$\gamma=4.3$ for $N=10^7$. From the early-$t$ behavior of
$\left<R_{ee}\right>$ (or the data for $t \lesssim 10$ in
\FIG{fignu}(a)) the obtained value of the exponent $\nu$ for the
relation
\bea%
\label{nu}%
\left<R_{ee}(t)\right> \sim t^\nu %
\eea
is $\nu\simeq 0.90(5)$.
The inset in \FIG{fignu}(a) displays the dependence of the
measured $\nu$ on the network size $N$, which indicates that the
values of $\nu$ approach to 1 as increasing $N$. This result
agrees with the expectation on the SWN \cite{almaas} when $p
\rightarrow 1$. The $\left<R_{ee}(t)\right>$ does not increase
indefinitely, but rapidly reaches a saturation value
$\left<R_{ee}({\bar \ell},t\rightarrow \infty)\right>$ after a
very short crossover time $\tau_{ee}$. Since $\tau_{ee}$ is nearly
equal to 10 or slightly larger than 10 even for the very large
network size (or $N=10^7$), the expectation of $\nu=1$ from the
data for nearly one decade of $t$ or so seems to be physically
dangerous. However, since an earlier analytical study on the walks
on a Cayley tree also suggests such $\nu=1$ behavior
\cite{yupcay}, $\nu \rightarrow 1$ as $N \rightarrow \infty$ is
physically plausible one. We have also measured $\nu$ for the size
upto $N=10^7$ and obtained the similar data as in \FIG{fignu}(a).
We thus expect that $\nu \rightarrow 1$ as $N \rightarrow \infty$
regardless of $\gamma$.

In \FIG{fignu}(b) we display the $\left<R_{ee}({\bar
\ell},t\rightarrow \infty)\right>$ for $\gamma=4.3$ as a function
of $\bar \ell$ with $\bar\ell\sim \ln N$. As shown in
\FIG{fignu}(b) $\left<R_{ee}({\bar \ell},t\rightarrow
\infty)\right>$ for $\gamma=4.3$ satisfies the power-law:
\bea %
\left<R_{ee}({\bar \ell},t\rightarrow \infty)\right> \sim {\bar
\ell}^\alpha. %
\label{alpha}%
\eea%
with $\alpha=0.93(2)$. $\left<R_{ee}({\bar \ell},t\rightarrow
\infty)\right>$ for various $\gamma$ is also found to satisfy the
power law (\ref{alpha}) very well. The obtained $\alpha$'s for
various $\gamma$ by assuming $\bar\ell\sim \ln\ln N$ for
$\gamma<3$ and $\bar\ell\sim \ln N$ for $\gamma\ge3$ are displayed
in Table \ref{tab:table1}.

\begin{figure}
\includegraphics[width=8.5cm]{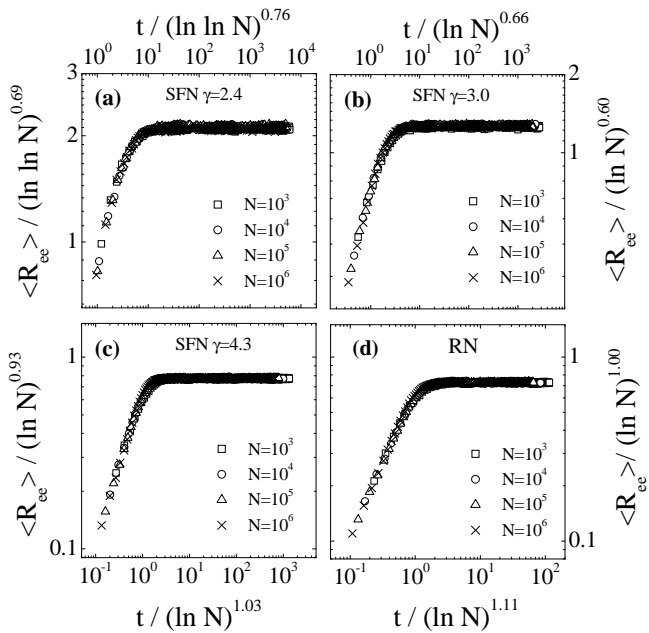}
\caption{Scaling plot of $\left<R_{ee}\right>$ on SFNs with (a)
$\gamma=2.4$ , (b) $3.0$, (c) $4.3$ and on (d) RNs  for $N=10^3,
10^4, 10^5$ and $10^6$. } \label{r2srw}
\end{figure}

%
From the Eqs. (\ref{nu}) and (\ref{alpha}), the dynamical scaling
relation
\begin{equation} \label{r2ansatz}
\left<R_{ee}({\bar \ell},t)\right>= \overline{\ell}^\alpha (\gamma
, N) g(t/\overline{\ell}^z ).
\end{equation}
%
is expected. The scaling function then $g(x)$ satisfies the
relation
\begin{equation}\label{r2func}
g(x)\sim \left\{
\begin{array}{lll}
x^{\nu}& , & x \ll 1\\
\mbox{const.} & , & x \gg 1.
\end{array}\right.
\end{equation}
The dynamical scaling relation (\ref{r2ansatz}) also physically
means that $\tau_{ee}$ scales as $\tau_{ee} \sim {\bar \ell}^z$.
The dynamic exponent $z$ can thus be evaluated from the measured
$\alpha$ and $z$ through the relation $z=\alpha / \nu$ (see Table
\ref{tab:table1}).
%
%
%
\FIG{r2srw} shows the scaling plot of $\left<R_{ee}\right>$
measured on SFNs with $\gamma=2.4$, $3.0$, $4.3$ and on random
networks (RNs) for $N=10^3,10^4, 10^5$ and $10^6$ using
\EQ{r2ansatz}. As shown in \FIG{r2srw}, $\left<R_{ee}\right>$'s
for various $N$ collapse very well into a single scaling curve
with the exponents listed in Table \ref{tab:table1} for each
network topology. Note that $\overline{\ell}$ in networks
represents the average diameter by definition \cite{DM}. From the
scaling collapse, we find the dependency of $\overline{\ell}$ on
$\gamma$ and $N$:
\begin{equation}
\label{elld} \overline{\ell} (\gamma,N) \sim d \sim \left\{
\begin{array}{lll}
\ln \ln N& , & 2<\gamma < 3\\
\ln N& , & \gamma \geq 3.\\

\end{array}\right.
\end{equation}
%

\begin{figure}
\includegraphics[width=7cm]{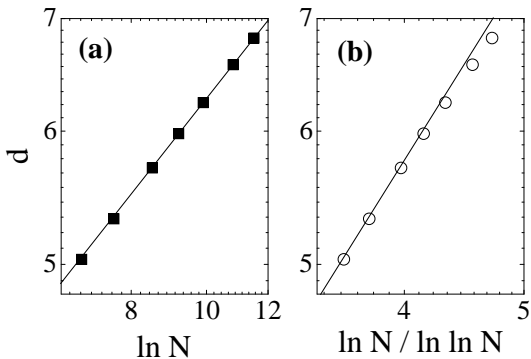}
\caption{Dependence of the diameter of static SFN with $\gamma=3$
on $N$. (a) The relation $d\sim \ln N$ is well satisfied. (b) The
relation $d \sim \ln N / \ln \ln N$ deviates when $N$ increases.}
\label{dia}
\end{figure}

For SFN with $\gamma=3$, it is known that the diameter scales as
$d\sim\ln N/(\ln \ln N)$ in the Barab{\'a}si-Albert (BA) model
\cite{basci} with looped structures \cite{handbook_born, szabo}.
In contrast the BA network without looped structures or tree-type
BA network satisfies $d\sim \ln N$ \cite{handbook_born, szabo}.
For the best scaling collapse we use the relation $d\sim \ln N$
for $\gamma=3$ even though the static SFN \cite{SNU_staticmodel}
has loops (see Fig. \ref{r2srw}(b)). To check the scaling behavior
$d \sim \ln N$ for $\gamma= 3$ obtained by RW method, the diameter
of SFN for $\gamma=3$ is directly measured by the BFA method (see
\FIG{dia}).
As can be seen from Fig. \ref{dia}(a), the relation $d \sim \ln N$
is well satisfied for the static SFN with $\gamma=3$ and agrees
with the result obtained by RW method or from Fig. \ref{r2srw}.
However, the relation $d \sim \ln N / \ln \ln N$ deviates when $N$
increases (Fig. \ref{dia}(b)).

The results from Eqs. (6) and (7) provide another way to find the
scaling behavior of $d$ in SFN. The result implies that the
computing time needed for the measurement of scaling behavior of
$d$ by RW method increases as $O((\ln N)^z)$ or $O((\ln \ln
N)^z)$. Since the computing time to measure $d$ by BFA increases
as $O(N^2)$, the RW method is far more efficient to find scaling
behavior of $d$.

\begin{table}
\caption{\label{tab:table1} Estimates of the exponents $\alpha$,
$z$ for SFNs with $\gamma$ and random network (RN). ($z=\alpha /
\nu$)}
\begin{ruledtabular}
\begin{tabular}{rrr|rrr}
$\gamma$&$\alpha$&$z$&$\gamma$&$\alpha$&$z$\\
\hline
2.15&0.42(1)&0.46(3)&3.0&0.60(1)&0.66(4)\\
2.40&0.69(2)&0.76(3)&3.5&0.81(1)&0.90(4)\\
2.50&0.84(1)&0.93(4)&4.3&0.93(2)&1.03(5)\\
2.75&1.12(2)&1.24(5)&5.7&0.98(4)&1.08(7)\\
2.90&1.49(1)&1.65(6)&&&\\
\hline&&&RN&1.00(5)&1.11(8)\\
\end{tabular}
\end{ruledtabular}
\end{table}
%

\begin{figure}
\includegraphics[width=8cm]{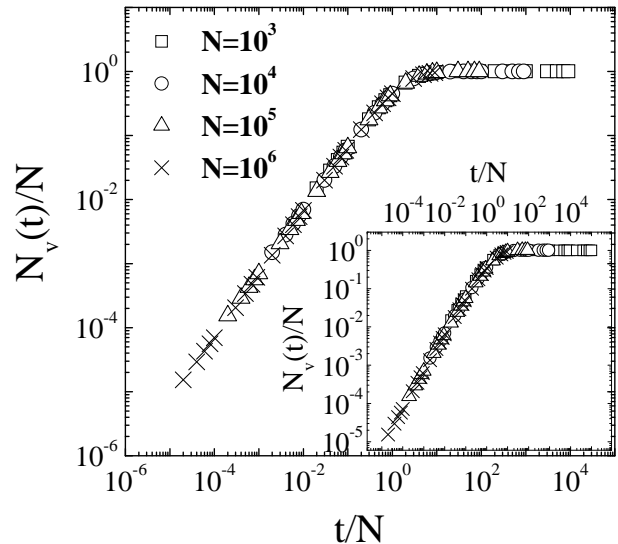}
\caption{This figure shows the scaling collapse for $N_v$ on SFN
with $\gamma=4.3$ and $2.4$ (inset).} \label{nv}
\end{figure}

For comparison's sake we now explain the average number of
distinct visited sites $N_v(t)$ on SFNs with various $\gamma$.
Fig. \ref{nv} shows the scaling plot of $N_v(t)$ for $\gamma=4.3$
and $2.4$ against $t/N$ for $N=10^3,10^4,10^5$ and $10^6$.
$N_v(t)$ on SFN satisfies the scaling relation
\begin{equation} \label{nvansatz}
N_v = Nf(t/N)
\end{equation}
with
\begin{equation} \label{nvfunc}
f(x)\sim \left\{
\begin{array}{lll}
x & , & x \ll 1\\
\mbox{1} & , & x \gg 1.
\end{array}\right.
\end{equation}
We have checked that the scaling relation (\ref{nvansatz}) holds
very well for SFNs with any $\gamma$.

The scaling relation (\ref{nvansatz}) is slightly different from
that of the regular lattice and the SWN. On the infinite
$D$-dimensional lattices $N_v$ depends on $D$. In the limit of
$t\rightarrow \infty$ \cite{erdos30}, $N_v \sim \sqrt{t}$ in
$D=1$, $N_v \sim t/\ln{t}$ in $D=2$ and $N_v \sim t$ for $D > 2$.
On SWNs, Almaas et el. \cite{almaas} showed that $N_v \sim
\sqrt{t}$ for $t \ll \xi^2$ (with $\xi=1/p$), $N_v \sim t$ for
$\xi^2 \ll t \ll N\xi$ and $N_v =N$ for $t \gg N\xi$. This means
that if the random walker does not reach the shortcut, then the
walker always sees the regular lattice structure ($D=1$). However,
if the walker meets the shortcut, then the behavior of walker
follows the mean-field result and finally $N_v$ saturates to $N$
due to the finite-size of the network. However, each node in SFN
can be connected to any other nodes and SFN can be regarded as an
infinite dimensional structure. Thus, $N_v$ on SFN follows the
mean-field behavior after the first few steps. From Eqs.
(\ref{nvansatz}) and (\ref{nvfunc}) the time $t_\times$ at which
$N_v=N$ is given by $t_\times\simeq N$. This implies that the
walker can sample a new region at each time step by using
shortcuts until it visits all nodes in the network. Thus the
statistical properties such as the average of certain quantity can
be significantly enhanced even for small RW steps.

Moreover, since $\left<R_{ee}(t)\right>\sim t^{1/2}$ and
$\left<N_v(t)\right>\sim t$ for $D\ge 2$, we can obtain a relation
between the $N_v$ and the number of accessible nodes (or volume)
$N_{ac}$ within the radius $\left<R_{ee}(t)\right>$ for $D\ge 2$:
\bea %
\label{vv}%
{N_v \over N_{ac}}\sim t^{1-D/2}.
\eea %
Thus, $N_{ac}/N_v$ diverges as $t\rightarrow \infty$ which
indicates the {\it{transient}} behavior of RWs \cite{Polya}. The
transient behavior becomes much more striking in the limit
$D\rightarrow \infty$. From $\tau_{ee} \sim (\ln N)^z$ (or ($\ln
\ln N)^z$ and $t_\times \simeq N$, we know $\tau_{ee} \ll
t_\times$ on SFN. This result means that the walker effectively
moves to the end of SFN without visiting all nodes. As a result,
we can measure the scaling behavior of $d(N)$ in much shorter time
than BFA algorithm (for example, see Figs. \ref{r2srw} and
\ref{dia}).

We investigate the scaling properties of RWs on SFNs. We measure
the end-to-end distance $R_{ee}$ and the number of distinct
visited sites $N_v$. From the scaling ansatz for
$\left<R_{ee}\right>$ as Eq. (\ref{r2ansatz}), we have measured
the scaling exponents $\alpha$, $\nu$ and $z$ for
$\left<R_{ee}\right>$ on various networks. From the scaling
relation we find the dependency of $\overline{\ell}$ on $\gamma$
and $N$.
Based on the simple scaling arguments, we also discuss the
theoretical reasons why scaling behavior of $d(N)$ of networks can
be measured far more efficiently by RW method than by BFA.

We thank to Dr. Kwon and Dr. Yoon for useful discussions. This
work is supported by grant No. R01-2006-000-10470-0 from the Basic
Research Program of the Korea Science \& Engineering Foundation.

\end{document}